\documentclass{ws-ijmpd}
\usepackage[super]{cite}
\usepackage{xcolor}
\usepackage[verbose,hypertexnames=false]{hyperref}
\hypersetup{colorlinks=false,allbordercolors=blue,pdfborderstyle={/S/U/W 1}}

\begin{document}

\markboth{Vindhyawasini Prasad}
{Instructions for Typing Manuscripts (Paper's Title)}

%
\catchline{}{}{}{}{}
%

\title{New Physics Searches at BESIII}

\author{Vindhyawasini Prasad}

\address{College of Physics, Jilin University\\
Changchun 130012, People’s Republic of China, China \\
vindy@jlu.edu.cn}

\maketitle

\begin{history}
\received{(Day Month Year)}
\revised{(Day Month Year)}
\accepted{(Day Month Year)}
\published{(Day Month Year)}
\end{history}

\begin{abstract}
This report highlights the recent BESIII results related to the new physics searches motivated by the shortcomings of Standard Model, such as the exclusion  of dark matter (DM). DM has so far been inferred only through astrophysical observations and accounts for a large fraction of matter density of the universe. DM may couple to SM particles via various portals. These portals give rise to several possible new particles beyond the SM, such as light Higgs bosons, dark photons, axion-like particles, or spin-1/2 fermions. Furthermore, the origin of DM and the observed asymmetry between visible matter and antimatter may be connected through the introduction of a dark baryon. The signature of these new physics particles could be accessible by the BESIII if their masses lie in the few-GeV range.

\end{abstract}

\keywords{Standard Model; Dark Matter; BESIII.}

\section{Introduction}	
The Standard Model (SM) describes spin-1/2 quarks and leptons, their fundamental interactions~\refcite{SM,GIM}, and the mechanism by which they acquire their masses~\cite{Higgs, Higgs-discovery}. However, it is not a complete theory of our universe. For example,  cosmological phenomena like galaxy rotations or nebula collisions can not be explained by the SM, but  infer the existence of dark matter and dark energy~\cite{Planck}. An extension of the SM is desirable to include the missing description of DM~\cite{DM}. DM amounts to about $27\%$ of the total matter density of the universe and is completely invisible across the entire electromagnetic spectrum. It has so far been inferred only through its gravitational effects on visible matter.

A sub-GeV scale particle carrying baryon number may explain the puzzle of the baryon asymmetry of the universe~\cite{baryogenesis}. A non-massive dark gauge boson may also enhance the decay rate of flavor-changing neutral current (FCNC) processes~\cite{JusakTand}, which are highly suppressed in the charm sector due to the Glashow–Iliopoulos–Maiani (GIM) mechanism~\cite{GIM}, making them accessible to the current generation of collider experiments. Despite many astrophysical observations of DM~\cite{Vera,anomalies}, it has not yet been observed in particle physics experiments. Many extensions of the SM, such as the \rm{\lq dark hidden sector\rq} model, open the possibility of coupling DM to SM particles via portals~\cite{Essig}. These portals introduce several new physics particles, such as light Higgs bosons, axion-like particles (ALPs), dark photons ($\gamma'$), or spin-1/2 sterile neutrinos. If the masses of these particles lie in the MeV–GeV range, they can be accessible at $e^+e^-$ collider experiments such as BESIII~\cite{bes3}.

BESIII is a China-based symmetric $e^+e^-$ collider experiment running in the tau-charm energy region~\cite{bes3}. It has explored several DM scenarios using data sets collected at various energy points, including the $J/\psi$, $\psi(3686)$, and $\psi(3770)$ resonances. Recent BESIII results related to the new physics searches include searches for a light Higgs boson~\cite{lightHiggs}, an ALP~\cite{ALP}, massive~\cite{massiveDP} and massless~\cite{masslessDP, ctougamma, Sigma} dark photons, sub-GeV dark gauge boson~\cite{subGeV} and baryon~\cite{DB},  muon-philic vector or scalar boson~\cite{muonphilic}, invisible decays of the $K_S$ meson~\cite{InvisKs}, and $\Lambda$--$\bar{\Lambda}$ oscillation~\cite{LambdaOsc}, as discussed in Ref.~\cite{Vindy}. This paper reviews the recent results of the BESIII experiment on new physics searches.

\section{Search for light gauge boson via radiative \boldmath{$J/\psi$} decay}
\begin{figure*}
\includegraphics[width=1.0\textwidth]{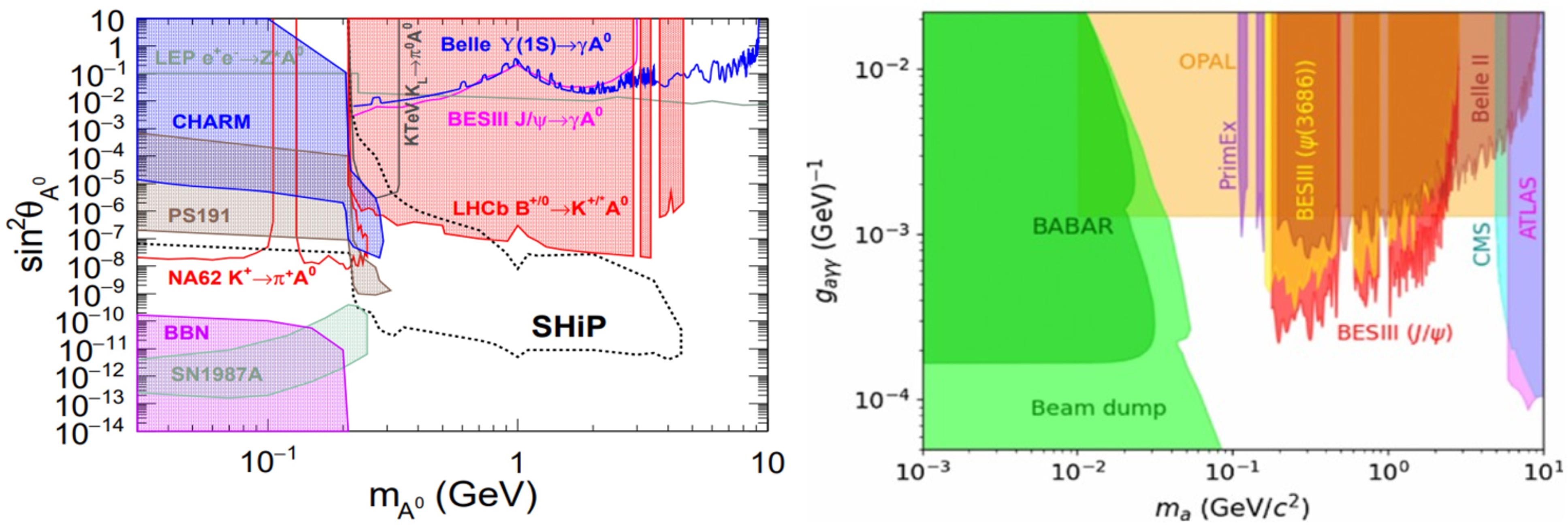}
\caption{\label{radGauge} 
(Left) $90\%$ CL upper limits on the mixing angle $\sin^2\theta_{A^0}$ as a function of the $A^0$ mass. 
(Right) $95\%$ CL upper limits on the axion–photon coupling $g_{a\gamma\gamma}$ as a function of the ALP mass $m_a$.}
\end{figure*}

The radiative decays of $J/\psi$ mesons provide a golden avenue to search for new physics beyond the SM~\cite{Wilczek}, such as light Higgs boson~\cite{lightHiggs} and ALP~\cite{ALP}. The light Higgs boson is predicted by many models beyond the SM, such as the Next-to-Minimal Supersymmetric Standard Model (NMSSM)~\cite{Maniatis}. The Higgs sector of the NMSSM contains seven Higgs bosons, among which there is a light $CP$-odd Higgs boson ($A^0$) that could be accessible via radiative decays of $J/\psi$~\cite{Wilczek}. The effective Yukawa coupling of the Higgs field to quark-pairs is proportional to $\cot\beta$ for up-type quarks and $\tan \beta$ for down-type quarks, where $\tan\beta$ is the ratio of the vacuum expectation values of the up- and down-type Higgs doublets. The branching fraction of $J/\psi \to \gamma A^0$ is expected to vary from $10^{-9}$ to $10^{-7}$, depending on the $A^0$ mass, $\tan\beta$, and NMSSM parameters~\cite{Dermisek}. The $A^0$ has been explored by various collider experiments, including BaBar~\cite{BaBar}, Belle~\cite{Belle}, and BESIII~\cite{lightHiggs}, but so far only null results have been reported. BESIII has recently reported $90\%$ confidence level (CL) upper limits on the product branching fraction $\mathcal{B}(J/\psi \to \gamma A^0) \times \mathcal{B}(A^0 \to \mu^+\mu^-)$ and on the effective Yukawa coupling of the Higgs field to charm quark pairs. The new BESIII results are slightly better than those of the BaBar~\cite{BaBar} and Belle~\cite{Belle} measurements, as seen in Fig.~\ref{radGauge} (left).

ALPs are pseudoscalar particles that arise in extensions of the SM and are inspired by the QCD axion, which was originally introduced through the spontaneous breaking of Peccei–Quinn symmetry~\cite{Peccei-Quinn} to solve the strong $CP$ problem in QCD~\cite{Strong-CP}. They are also proposed as cold DM candidates. BESIII has recently performed a search for ALPs decaying into two photons via radiative decays of $J/\psi$. No significant signal events were observed, and $95\%$ CL upper limits were set on the product branching fraction $\mathcal{B}(J/\psi \to \gamma a) \times \mathcal{B}(a \to \gamma \gamma)$ and on the axion–photon coupling ($g_{a\gamma\gamma}$) in the mass range of $[0.18,,2.85]$ GeV/$c^2$. The obtained BESIII limits on $g_{a\gamma\gamma}$ are the most stringent to date~\cite{ALP}, as seen in Fig.~\ref{radGauge} (right).

\section{New physics signature with invisible final state}
The invisible final states of light mesons may probe various flavors of new physics particles, such as massless dark photons~\cite{masslessDP, ctougamma, Sigma}, sub-GeV DM particles~\cite{subGeV} , and massive muon-philic scalar or vector bosons~\cite{muonphilic}. The following sub-sections review these new physics scenarios with invisible final states.

\subsection{Search for massless dark photon via $D^0 \to \omega \gamma'$ and $D^0 \to \gamma \gamma'$}
\begin{figure*}
\includegraphics[width=1.0\textwidth]{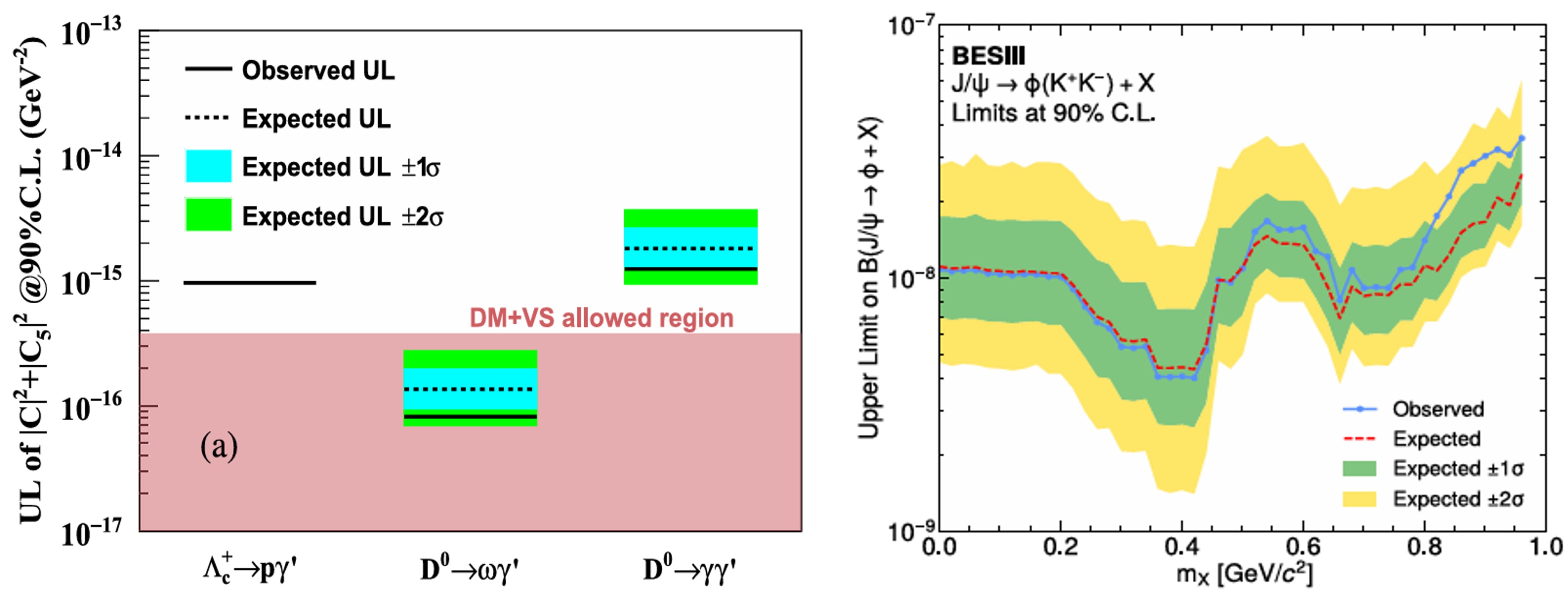}
\caption{\label{Invis} 
(Left) The $90\%$ CL upper limits on the effective coupling operator $|C|^2 + |C_5|^2$ obtained from the explored decay mode in massless dark photon searches. (Right) The $90\%$ CL upper limits on $\mathcal{B}(J/\psi \to \phi +X)$ as a function of sub-GeV invisible particle $X$ mass. }
\end{figure*}

A simple extension of the SM accommodates an extra Abelian $U(1)_D$ symmetry that remains unbroken, introducing a massless dark photon ($\gamma'$)~\cite{dsix}. The presence of a massless dark photon may enhance the decay rate of FCNC processes, making them accessible with existing collider experiments~\cite{JusakTand}. The search for a massless dark photon at BESIII is based on 7.9 fb$^{-1}$ of quantum-correlated $\psi(3770) \to D^0 \bar{D}^0$ data using a double-tag technique~\cite{ctougamma}. One of the $\bar{D}^0$ candidates is tagged using its dominant hadronic decay modes, while the other $D^0$ meson is reconstructed in the signal mode of interest. The signal of a massless dark photon is inferred from a fit to the distribution of the missing mass squared. No evidence for massless dark photon production is found, and $90\%$ CL upper limits are set at $\mathcal{B}(D^0 \to \omega \gamma') < 1.1 \times 10^{-5}$ and $\mathcal{B}(D^0 \to \gamma \gamma') < 2.0 \times 10^{-6}$. The obtained result from $D^0 \to \omega \gamma'$ improves the constraint on the effective coupling operator $|C|^2 + |C_5|^2 < 8.2 \times 10^{-17}$ GeV$^{-2}$, reaching regions allowed by dark matter and vacuum stability considerations and surpassing previous $\Lambda^+ \to p \gamma'$ limits by an order of magnitude~\cite{ctougamma}, as seen in Fig.~\ref{Invis} (left).

\subsection{Search for sub-GeV invisible particle in $J/\psi \to \phi X$}
Sub-GeV light DM candidates behave as invisible particles due to their weak coupling to SM particles~\cite{subGeV1}. Such a DM particle could be an ALP~\cite{ALP} or dark photon~\cite{massiveDP} and be accessible via $J/\psi \to \phi + \mathrm{invisible}$. The search for an invisible particle, $X$, with a mass between 0 and 0.96 GeV/$c^2$ has been performed in the process $J/\psi \to \phi X$ using 8.8 billion $J/\psi$ events collected by the BESIII experiment~\cite{subGeV}. The search is performed in the mass range from 0 to 0.96 GeV/$c^2$ to avoid the large background from $J/\psi \to \phi K_L^0 K_L^0$. No evidence of significant signal events is found, and $90\%$ CL upper limits on the branching fraction are set in the range of $(4$–$40) \times 10^{-9}$ over the investigated mass range. In addition, the $90\%$ CL upper limit on the branching fraction of $\eta \to \mathrm{invisible}$ is determined to be $2.4 \times 10^{-5}$, which has four times improvement over the previous measurements~\cite{etameasurement}.

\subsection{Search for a dark baryon in the $\Xi^- \to \pi^- + {\rm invisible}$ decay}
\begin{figure*}
\includegraphics[width=1.0\textwidth]{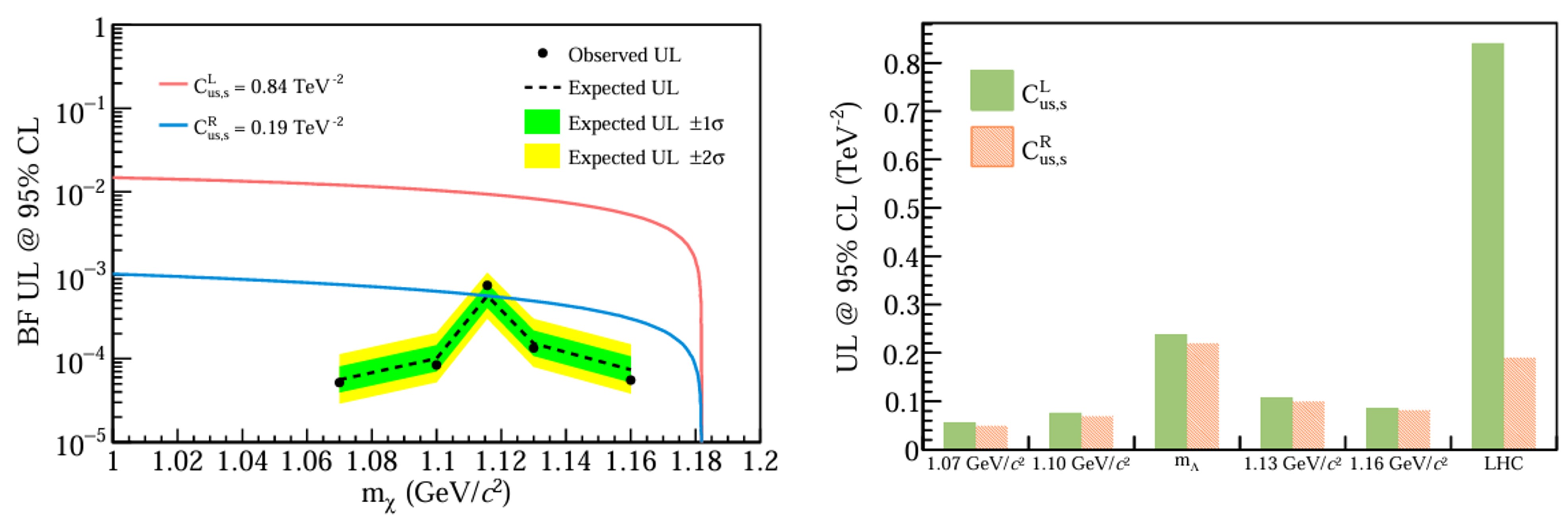}
\caption{\label{DB} (Left) The $95\%$ CL upper limits on the expected and observed branching fractions of $\Xi^- \to \pi^- +{\rm invisible}$ as a function of dark baryon mass. (Right) The $95\%$ CL upper limits on the Wilson coefficients $C_{us,s}^L$ and $C_{us,s}^{R}$ derived from the results under different mass hypotheses, together with those from LHC searches. }
\end{figure*}

Dark matter's density suggests a link to baryonic matter, potentially carrying baryon number~\cite{rhoDM}. This connection is explained by mechanisms like $B$-mesogenesis, which could solve both the dark matter origin and matter-antimatter asymmetry~\cite{mesogenesis}.  More recently, BESIII performed a search for a dark baryon $\chi$ in the two-body decay $\Xi^- \to \pi^- + \rm{invisible}$ using approximately $10^7$ events from $J/\psi \to \Xi^- \bar{\Xi}^+$ decays, based on 10 billion $J/\psi$ events~\cite{DB}.  No evidence of dark baryon production is found, and $90\%$ ($95\%$) CL upper limits on the branching fraction $\mathcal{B}(\Xi \to \pi^- + {\rm invisible})$ are set as a function of the $\chi$ mass, as shown in Fig.~\ref{DB} (left). The $95\%$ CL upper limits on Wilson coefficients $C_{us,s}^L$ and $C_{us,s}^R$ are set to be less than $5.5 \times 10^{-2}$ TeV$^{-2}$ and $4.9 \times 10^{-2}$ TeV$^{-2}$, respectively, which are more stringent than the previous limits from LHC experiments~\cite{DB}, as seen in Fig.~\ref{DB} (right).

\subsection{Search for muon-philic scalar or vector particle in $J/\psi \to \mu^+\mu^- {\rm invisible}$}
A new type of massive vector boson $X_1$ or scalar $X_0$ may appear in an extended SM sector with anomaly-free gauge symmetries $U(1)$ or $U(1){L{\mu}-L_{\tau}}$~\cite{muphilic}. These particles couple only to the second or third generation of leptons with coupling strength $g'{0,1}$, originally proposed to explain the long-standing $(g-2)\mu$ anomaly, and may also explain the dark matter abundance. Searches for these $X_{0,1}$ particles have been performed via $J/\psi \to \mu^+\mu^- X_{0,1}$ with $X_{0,1}$ decaying invisibly, using 9 billion $J/\psi$ events~\cite{muonphilic}. No evidence of significant signal events is found, and $90\%$ CL upper limits are set on $g'{0,1}$ in the vanilla $L{\mu}-L_{\tau}$ model, the invisible $L_{\mu}-L_{\tau}$ model, and the scalar $U(1)$ model, as detailed in Ref.~\cite{muonphilic}.

\subsection{Invisible $K_S^0$ decays}
In the SM, the branching fraction of the FCNC decay $K_S^0 \to \nu \bar{\nu}$ is highly suppressed due to angular momentum conservation and helicity suppression, resulting in a predicted branching fraction of less than $10^{-16}$~\cite{Kstoinvis}. Contributions from new physics models, such as the two-Higgs-doublet model~\cite{Kstoinvis} and mirror-matter model~\cite{mirror}, could enhance the branching fraction of the $K_S^0$ invisible decay up to the level of $10^{-6}$. Therefore, a search for invisible decays of $K_S^0$ not only probes new physics but also provides valuable input for tests of charge, parity, and time-reversal (CPT) symmetry through the Bell–Steinberger relation, which connects CPT violation to the amplitudes of all decay channels of neutral kaons under the assumption that no invisible decay modes exist~\cite{InvisKs}. BESIII performed a search for invisible decays of the $K_S^0$ meson using 10 billion $J/\psi$ events via the decay $J/\psi \to \phi K_S^0 K_S^0$, where one $K_S^0$ is tagged through its $K_S^0 \to \pi^+\pi^-$ decay mode and the other $K_S^0$ meson is allowed to decay invisibly~\cite{InvisKs}. The $J/\psi \to \phi K_S^0 K_S^0$ decay with both $K_S^0$ mesons decaying into pion pairs is also studied as a normalization channel to reduce possible sources of systematic uncertainty. No evidence of a significant signal is found, and a $90\%$ CL upper limit on $\mathcal{B}(K_S^0 \to \mathrm{invisible})$ is set to be less than $8.4 \times 10^{-4}$ for the first time~\cite{InvisKs}.

\section{Search for $\Lambda-\bar{\Lambda}$ osicallation}
Though the SM conserves baryon number, the universe still exhibits a dominant matter asymmetry. This asymmetry can be understood in terms of the three Sakharov conditions: the existence of charge (C) and parity (P), or charge–parity (CP), violation; baryon-number-violating interactions; and the existence of out-of-thermal-equilibrium conditions in the early universe. According to grand unified theories, protons can decay into light quarks in a variety of ways~\cite{Sakharov}.  Some grand unified theories predict that protons can decay via baryon number violating processes in the final state~\cite{BNV}.  This mechanism simultaneously breaks the conservation of baryon number ($B$) and lepton number ($L$) while keeping their difference, $B-L$, conserved~\cite{pdg}. If neutrinos are Majorana particles with small masses~\cite{Majorana}, this would imply the presence of $\Delta(B-L)=2$ interactions, thereby suggesting the existence of nucleon–antinucleon ($n$–$\bar{n}$)~\cite{nnbar} and $\Lambda$–$\bar{\Lambda}$~\cite{lambdalambdabar}  oscillations. If $\Lambda$–$\bar{\Lambda}$ oscillation occurs, the time-integrated probability of $\Lambda$–$\bar{\Lambda}$ conversion is given by:
\begin{equation}
P(\Lambda)=\frac{\int_{0}^{\infty}\sin^{2}(\delta m_{\Lambda\bar{\Lambda}}t)e^{-t/\tau_{\Lambda}}dt}
{\int_{0}^{\infty}e^{-t/\tau_{\Lambda}}dt}
\end{equation} 

\noindent where $\delta m_{\Lambda\bar{\Lambda}}$ is the transition mass between the $\Lambda$ and $\bar{\Lambda}$ in the oscillation at time $t$, and $t_{\Lambda} = (2.632 \pm 0.02) \times 10^{-10}$ s is the mean lifetime of the $\Lambda$. The oscillation parameter can be calculated as:
\begin{equation}
(\delta m_{\Lambda\bar{\Lambda}})^{2}=\frac{P(\Lambda)}{2(\tau_{\Lambda}/\hbar)^{2}}
\end{equation}
\noindent where $\hbar$ is the Planck constant. BESIII has recently reported an improved null result in the search for baryon-number violation via $\Lambda$–$\bar{\Lambda}$ oscillation in the decay $J/\psi \to \Lambda \bar{\Lambda}$ using 10 billion $J/\psi$ events~\cite{LambdaOsc}. The upper limits at $90\%$ CL are established for the time-integrated probability of $\Lambda$–$\bar{\Lambda}$ oscillation and the oscillation parameter, set at $P(\Lambda) < 1.4 \times 10^{-6}$ and $\delta m_{\Lambda \bar{\Lambda}} < 2.1 \times 10^{-18}$ GeV, respectively.

\section{Summary}
Searches for new physics beyond the SM are one of the prime objectives of the current generation of collider experiments. BESIII plays a unique role in searching for various flavors of dark matter candidates. While no positive signals have been observed to date, the stringent upper limits set by BESIII have significantly constrained the parameter space of various new physics models. These results complement searches at high-energy colliders by probing the low-mass region with high precision.  BESIII is expected to produce new results in the near future, especially with the recently collected 20 fb$^{-1}$ of $\psi(3770)$ data~\cite{bes3}.

\section*{Acknowledgments}
The author wishes to acknowledge the organizers of the PGU-3 2025 conference for their wonderful hospitality and for providing a stimulating environment for physics discussions.

This work is supported by the Seed Funding of Jilin University.

\section*{ORCID}

\noindent Vindhyawasini Prasad - \url{https://orcid.org/0000-0001-7395-2318}

\end{document}